\documentclass[%
 reprint,
 superscriptaddress,
 amsmath,amssymb,
 aps,
 pra,
 floatfix,
 twocolumn,
 longbibliography
]{revtex4-2}

\newcommand{\bra}[1]{\langle #1\rvert}
\newcommand{\ket}[1]{\lvert #1\rangle}

\newcommand{\op}[2]{\ket{#1} \bra{#2}}

\newcommand{\pd}[1]{\frac{\partial #1}{\partial t}}
\newcommand{\rpd}[1]{\partial_t #1}

\DeclareMathOperator{\sgn}{sgn}

\usepackage{graphicx}
\usepackage{dcolumn}
\usepackage{bm}
\usepackage{hyperref}

\usepackage{tikz}
\usetikzlibrary{calc}

\usepackage{orcidlink}

\usepackage{amsmath}
\usepackage{upgreek}
\usepackage{physics}
\usepackage{isomath}
\usepackage{placeins}
\usepackage{comment}
\usepackage{mathrsfs}

\begin{document}


\title{Speeding Up Squeezing with a Periodically Driven Dicke Model}
\author{Jarrod T. Reilly\orcidlink{0000-0001-5410-089X}}
\affiliation{JILA, NIST, and Department of Physics, University of Colorado, 440 UCB, Boulder, CO 80309, USA}
\author{Simon B. J\"ager\orcidlink{0000-0002-2585-5246}}
\affiliation{Physics Department and Research Center OPTIMAS, University of Kaiserslautern-Landau, D-67663, Kaiserslautern, Germany}
\author{John Drew Wilson\orcidlink{0000-0001-6334-2460}}
\affiliation{JILA, NIST, and Department of Physics, University of Colorado, 440 UCB, Boulder, CO 80309, USA}
\author{John Cooper}
\affiliation{JILA, NIST, and Department of Physics, University of Colorado, 440 UCB, Boulder, CO 80309, USA}
\author{Sebastian Eggert\orcidlink{0000-0001-5864-0447}}
\affiliation{Physics Department and Research Center OPTIMAS, University of Kaiserslautern-Landau, D-67663, Kaiserslautern, Germany}
\author{Murray J. Holland\orcidlink{0000-0002-3778-1352}}
\affiliation{JILA, NIST, and Department of Physics, University of Colorado, 440 UCB, Boulder, CO 80309, USA}

\date{\today}


\begin{abstract}
We present a simple and effective method to create highly entangled spin states on a faster timescale than that of the commonly employed one-axis twisting (OAT) model.
We demonstrate that by periodically driving the Dicke Hamiltonian at a resonance frequency, the system effectively becomes a two-axis countertwisting Hamiltonian which is known to quickly create Heisenberg limit scaled entangled states.
For these states we show that simple quadrature measurements can saturate the ultimate precision limit for parameter estimation determined by the quantum Cram\'er-Rao bound.
An example experimental realization of the periodically driven scheme is discussed with the potential to quickly generate momentum entanglement in a recently described experimental vertical cavity system.
We analyze effects of collective dissipation in this vertical cavity system and find that our squeezing protocol can be more robust than the previous realization of OAT.
\end{abstract}

{
\let\clearpage\relax
\maketitle
}

\section{Introduction}
For centuries, advancements in precision measurements have continuously propelled the scientific community's understanding of the fundamental nature of reality. 
This inspired both the quantum revolution and Einstein's theories of relativity, with the frontier of each still advancing through the use of increasingly precise experiments~\cite{Davisson,Freedman,Clemence,Shapiro,Abbott}.
Current state-of-the-art precision measurements can detect a change of mirror distance of $10^{-3}$ of the proton's width in gravitational wave detectors~\cite{Abbott,Abbott2,Aasi} and have led to the development of atomic clocks with a fractional frequency uncertainty of $10^{-21}$~\cite{Bothwell}, among many other groundbreaking achievements~\cite{Taylor,Templier,Akiyama,Xue,Roussy,Abi,Albahri,Morel,Agazie}.

Most precision metrology experiments still operate at or above the standard quantum limit (SQL), which is the fundamental sensitivity threshold that arises from shot noise in measurements of uncorrelated quantum states. 
This limit on product states can be overcome through the use of entangled quantum states, and if this can be consistently utilized, it would revolutionize precision measurements with the potential to discover new physics.
Although there have been proof of principle experimental demonstrations of quantum entanglement, applications for a true sensing purpose have so far been limited~\cite{Tse,PedrozoPenafiel,Hosten,Robinson,Malia}.
For example, spin squeezing offers a promising platform to perform atomic clock experiments beyond the SQL, but often require a long squeezing time during which quantum correlations may be destroyed by decoherence. 

In this work, we propose an experimentally relevant scheme to realize spin squeezing in a short propagation time.
We show that driving the Dicke model~\cite{Dicke,Baumann,Dimer,Chitra,Reilly2,Jager2} at a parametric resonance leads to an effective two-axis countertwisting (TACT) Hamiltonian which can reach Heisenberg limited scaling in a shorter timescale than the commonly employed one-axis twisting (OAT) Hamiltonian.
While the TACT Hamiltonian has been studied theoretically~\cite{Kitagawa,Law,Andre,Huang,Liu2,Yukawa,Kajtoch,Opatrny,Wu,Kruse,Borregaard,Anders,Zhang,Hernandez}, it has so far been elusive to achieve experimentally.
We demonstrate how TACT may be realized in a current, state-of-the-art vertical cavity experiment~\cite{Greve,Luo,ActuallyJarrodsPaper2,Zhang2} by periodically modulating an injected field that drives the cavity.
We discuss how to make optimal use of the system's entanglement for phase estimation using a recent advance that uncovers a state's full metrological potential by diagonalizing the quantum Fisher information matrix (QFIM)~\cite{Reilly}.
We then perform a Bayesian phase reconstruction sequence where, remarkably, we find that simple quadrature measurements saturate the quantum Cram\'er-Rao bound (QCRB)~\cite{Pezze}.

The structure of our paper is as follows:
We begin in Sec.~\ref{Sec:Dicke} by discussing the general idea and performance of generating fast spin squeezing in the Dicke model using parametric driving. Section~\ref{Sec:Exp} puts this model into a physical context where we derive a master equation from a realizable setup in a matter-wave system.
In Sec.~\ref{Sec:Diss}, we discuss the effect of dissipation on the squeezing performance and in Sec.~\ref{Sec:Conclusion}, we conclude our results.

\section{Periodically Driven Dicke Model\label{Sec:Dicke}}
We consider $N$ atoms that are collectively coupled through a cavity field. 
The atoms have ground state $\ket{\downarrow}$ and excited state $\ket{\uparrow}$, and we define the collective raising and lowering operators $\hat{J}_+ = \sum_j \ket{\uparrow}_j \bra{\downarrow}_j = \hat{J}_-^{\dagger}$.
This system has an underlying $\mathrm{SU}(2)$ symmetry with basis operators $\hat{J}_x = (\hat{J}_+ + \hat{J}_-) / 2$, $\hat{J}_y = i (\hat{J}_- - \hat{J}_+) / 2$, and $\hat{J}_z = [\hat{J}_+, \hat{J}_-] / 2$, as well as the quadratic Casimir operator $\hat{J}^2 = \hat{J}_x^2 + \hat{J}_y^2 + \hat{J}_z^2$.
After eliminating the cavity in the dispersive regime, we consider dynamics governed by the time-dependant Dicke Hamiltonian~\cite{Kirton} 
\begin{equation} \label{H_Dicke}
    \hat{H} = \hbar \Delta \hat{J}_z + \hbar \chi \cos(\omega t) \hat{J}_x^2,
\end{equation}
where $\Delta$ is a detuning and $\chi$ scales the cavity-mediated nonlinearity. 
This Hamiltonian can model, for example, Raman transitions between hyperfine states using two time-dependent transverse fields~\cite{Dimer,Reilly2}. 
For now, we ignore cavity decay based on large cavity detuning, such that the dynamics are governed by the Schr\"odinger equation $\rpd \hat{\rho} = -i [ \hat{H}, \hat{\rho} ] / \hbar$ with density matrix $\hat{\rho}$.
We discuss the effects of non-negligible dissipation in the next section.

The nonlinearity in Eq.~\eqref{H_Dicke} creates an entangled state which can be used to sense a physical parameter with a quantum advantage.
To find the parameter $\Phi$ that the generated state is most sensitive to, one finds the maximum quantum Fisher information (QFI), $\lambda_{\mathrm{max}}$, by calculating the largest eigenvalue of the quantum Fisher information matrix (QFIM)~\cite{Reilly},
\begin{equation} \label{OptimalGenerator}
    \bm{\mathcal{F}} \vec{\mathcal{O}} = \lambda_{\mathrm{max}} \vec{\mathcal{O}},
\end{equation}
where the elements of the QFIM are given by~\cite{Liu} 
\begin{equation}\label{QFIM}
    \bm{\mathcal{F}}_{\mu \nu} = \sum_{i,j = 0; \varrho_i + \varrho_j \neq 0}^{\dim[\hat{\rho}] - 1} \frac{2 \Re \left[ \bra{\varrho_i} \left[ \hat{J}_{\mu}, \hat{\rho} \right] \ket{\varrho_j} \bra{\varrho_j} \left[ \hat{\rho}, \hat{J}_{\nu} \right] \ket{\varrho_i} \right]}{\varrho_i + \varrho_j},
\end{equation}
with $\mu, \nu \in \{ x, y, z \}$ and the spectral decomposition $\hat{\rho} = \sum_i \varrho_i \op{\varrho_i}{\varrho_i}$. This is the description of the QFIM for general mixed states which will be of special importance later when we discuss dissipation in Sec.~\ref{Sec:Diss}.
The eigenvector $\vec{\mathcal{O}}$ associated with this maximum eigenvalue corresponds to the optimal generator $\hat{\mathcal{G}}$ that encodes the optimal parameter~\cite{Reilly}, $\exp[-i \hat{\mathcal{G}} \Phi]$.
The QFI for unentangled states can reach the SQL, $\lambda_{\mathrm{max}} = N$, while entangled states can reach the Heisenberg limit (HL), $\lambda_{\mathrm{max}} = N^2$, which is the fundamental limit on sensing set by the Heisenberg uncertainty principle~\cite{Holland}.

\begin{figure}
    \centerline{\includegraphics[width=\linewidth]{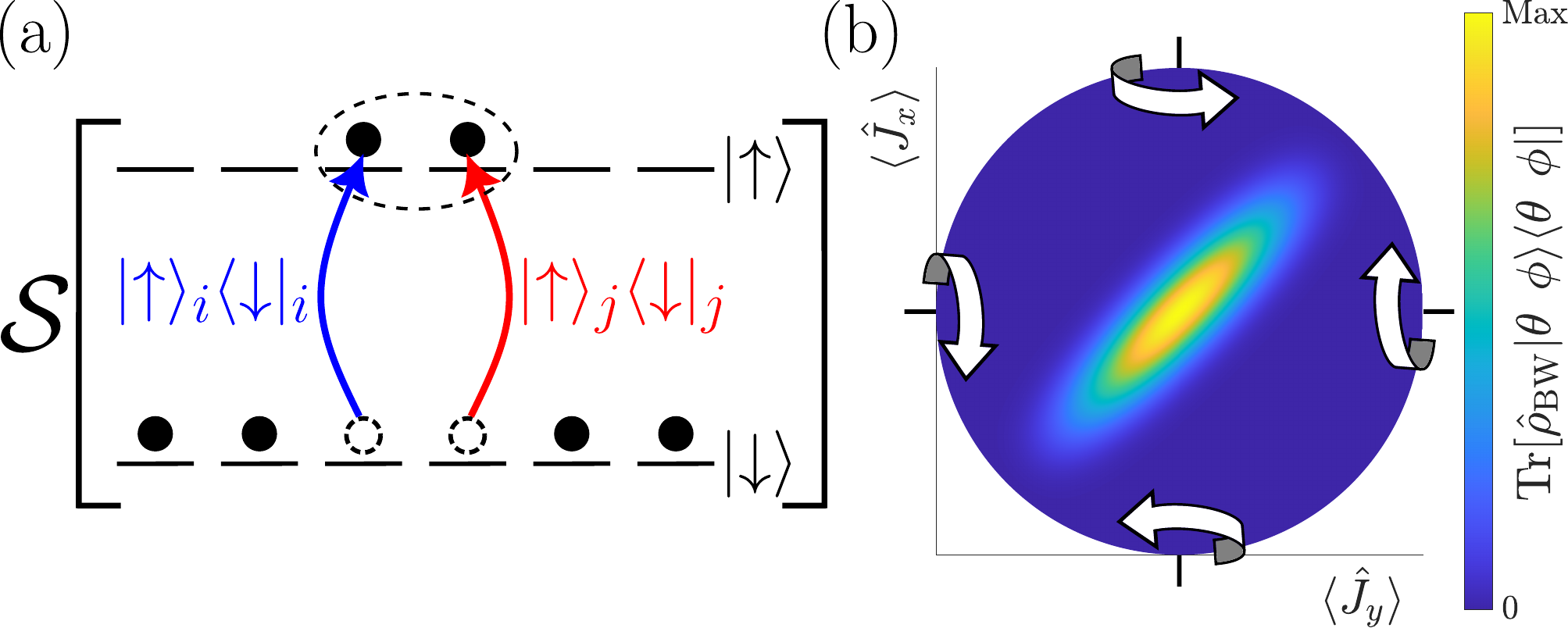}}
    \caption{(a) Schematic of the pair production process $\hat{J}_+^2$ that the PDD model drives (along with $\hat{J}_-^2$) to generate interparticle entanglement (dashed line).
    Here, $\mathcal{S}$ is the the symmeterizer which sums over all permutations of $i$ and $j$~\cite{Xu}.
    (b) The collective Bloch sphere for $\hat{\rho}_{\mathrm{BW}}$ in the rotating frame of Eq.~\eqref{H_tilde}.
    The color represents the state's overlap with the coherent spin state $\ket{\theta, \phi} = \exp[-i \phi \hat{J}_z] \exp[-i \theta \hat{J}_y] \ket{\downarrow}^{\otimes N}$ at each point.
    The arrows indicate the direction of twisting about each axis.}
    \label{fig:Schematic}
\end{figure}
\begin{figure}
    \centerline{\includegraphics[width=\linewidth]{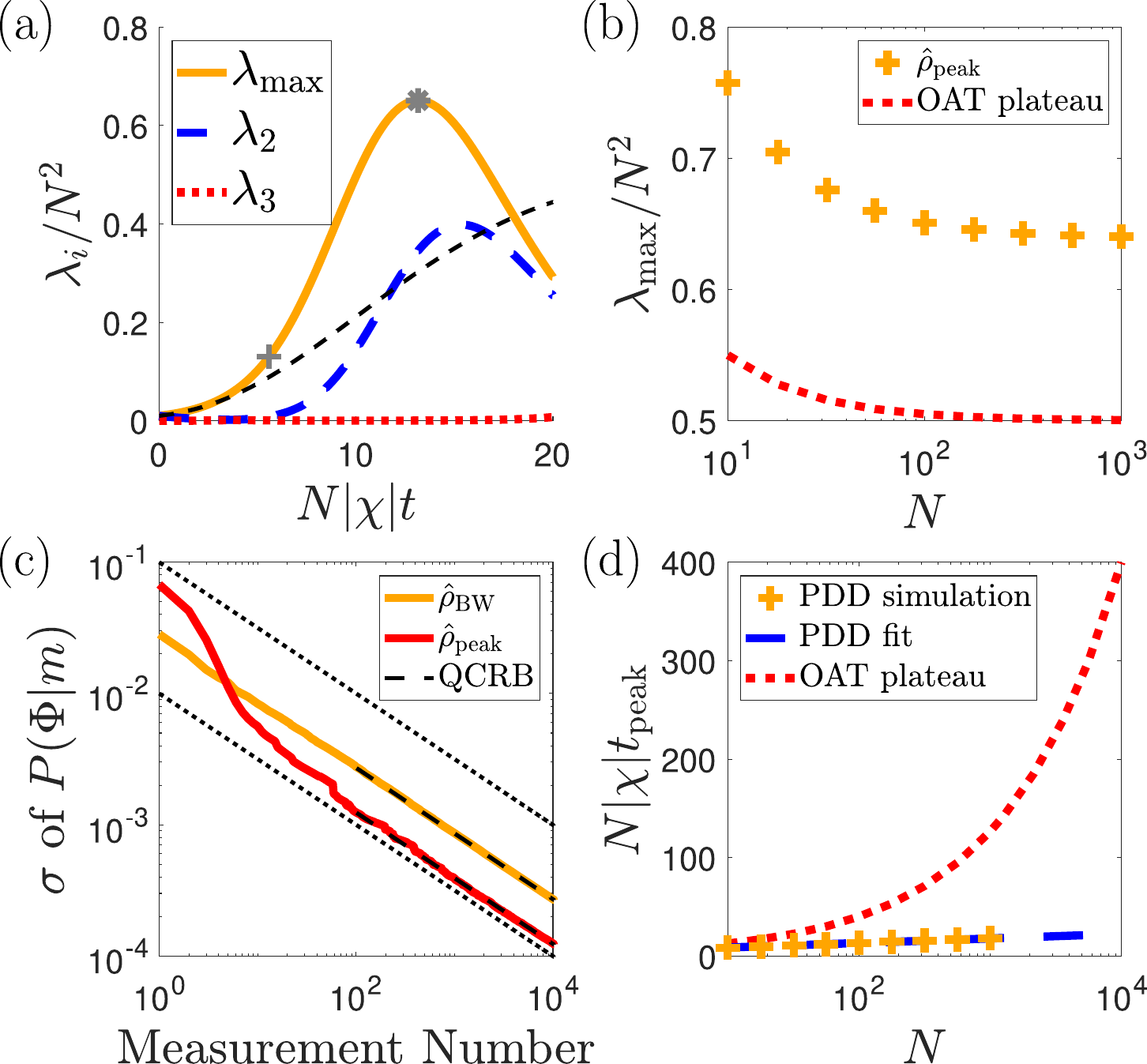}}

    \caption{(a) The three eigenvalues of the QFIM $\bm{\mathcal{F}}$ for $N = 100$.
    The two eigenvalues besides $\lambda_{\mathrm{max}}$ are labeled $\lambda_2$ and $\lambda_3$.
    The state evolves under Eq.~\eqref{H_Dicke} with $\Delta = 100 N \abs{\chi}$.
    The gray plus and asterisk indicate when the system reaches $\hat{\rho}_{\mathrm{BW}}$ and $\hat{\rho}_{\mathrm{peak}}$, respectively.
    Also shown is the largest eigenvalue of the QFIM for OAT with the same parameters (dashed black line).
    (b) The largest QFIM eigenvalue for $\hat{\rho}_{\mathrm{peak}}$.
    Also shown is the plateau value of $N (N + 1) / 2$ for OAT.
    (c) Sensitivity, given by the standard deviation $\sigma$ of the posterior distribution, for the optimal parameter $\Phi$ after applying Bayes theorem.
    We display results for the states $\hat{\rho}_{\mathrm{BW}}$ and $\hat{\rho}_{\mathrm{peak}}$, and the dashed lines represent the QCRB for the respective state.
    The top and bottom dotted lines represent the SQL and HL, respectively.
    (d) Comparing the time of maximum QFI for PDD $t_{\mathrm{peak}}$ (orange plus) with the time OAT reaches its plateau $t_{\mathrm{pl}}$ (dotted red line) for a constant $N \abs{\chi}$.
    We also show the curve fit of the PDD simulations given by Eq.~\eqref{t_peak} (dashed blue line).}
    \label{fig:QFIplot}
\end{figure}

Although we will use Eq.~\eqref{H_Dicke} for our numerical simulations, one can gain a better intuition of the dynamics by transforming into a rotating frame.
We move into an interaction picture $\tilde{\hat{\rho}} = \hat{U}^{\dagger} \hat{\rho} \hat{U}$ with $\hat{U} = \exp [ -i \Delta t\hat{J}_z]$, so that
Eq.~\eqref{H_Dicke} becomes
\begin{equation} \label{H_tilde}
    \tilde{\hat{H}} = \frac{\hbar \chi}{4} \cos(\omega t) \left[ e^{2 i \Delta t} \hat{J}_+^2 + 2 \left( \hat{J}_+ \hat{J}_- - \hat{J}_z \right) + e^{-2 i \Delta t} \hat{J}_-^2 \right].
\end{equation}
In the majority of previous works, one assumes a constant nonlinear interaction rate $\omega = 0$.
Then, in the limit $\abs{\Delta} \gg N \abs{\chi}$, one makes the rotating-wave approximation (RWA)~\cite{Steck} to drop the fast-oscillating $\hat{J}_{\pm}^2$ terms.
We now explore an opposite regime in which the system is instead driven on the parametric resonance $\omega = 2 \Delta$. 
In this case we label this as the periodically driven Dicke (PDD) model
\begin{equation} \label{H_PDD}
    \hat{H}_{\mathrm{PDD}}=\hbar \Delta \hat{J}_z + \hbar \chi \cos(2\Delta t) \hat{J}_x^2 .
\end{equation}
Equation~\eqref{H_PDD} after the RWA, which requires small $\chi\ll\Delta$, becomes 
\begin{equation} 
    \hat{H}_{\mathrm{PDD}} \approx \frac{\hbar \chi}{8} \left( \hat{J}_+^2 + \hat{J}_-^2 \right),
\end{equation}
which is seen by expanding $\cos(\omega t) = (\exp[i \omega t] + \exp[-i \omega t]) / 2$. 
This shows the connection of the PDD with two-axis countertwisting (TACT)~\cite{Kitagawa}, which was found to reach HL scaling on an exponential timescale~\cite{Andre,Kajtoch} through the pair production and twisting processes shown in Fig.~\ref{fig:Schematic}.
Beginning in the collective ground state $\hat{\rho}_0 = \op{\downarrow}{\downarrow}^{\otimes N}$, we examine the sensitivity of the PDD model using the maximum QFI from Eq.~\eqref{OptimalGenerator}.
We display the dynamics of the QFIM eigenvalues in Fig.~\ref{fig:QFIplot}(a) for the case of $N = 100$.
Here, one can see the exponential scaling of the maximum QFI on short timescales.
In the rotating frame of Eq.~\eqref{H_tilde}, we find that the optimal generator corresponding to $\lambda_{\mathrm{max}}$ is given by $\hat{\mathcal{G}} = ( \hat{J}_x + \hat{J}_y ) / \sqrt{2}$.
This can be understood by interpreting Eq.~\eqref{H_PDD} as an analog to the photonic Kerr nonlinearity~\cite{Caves} which can be formalized if one performs the Holstein-Primakoff approximation assuming low atomic excitations~\cite{Holstein,Byrnes}. 

During the initial squeezing, the state has a high overlap with the Berry-Wiseman (BW) phase state~\cite{Berry}, up to a rotation.
The BW phase state maximizes the information gained about an unknown phase after a single measurement~\cite{Berry2} and has a full $2 \pi$ dynamic range (see Appendix~\ref{BWappendix}).
For states generated by TACT, the fidelity with the phase state has been shown to reach unity~\cite{Kajtoch}. 
We find that the rapid-oscillating terms in the PDD model Eq.~\eqref{H_PDD} do not have a noticeable effect on this fidelity, and so the PDD model also reaches unit fidelity with the BW phase state up to a negligible error. 
For $N = 100$, this unit fidelity occurs at $t \approx 5.57 / (N \abs{\chi})$, and so we label this state $\hat{\rho}_{\mathrm{BW}}$ which we display on the collective Bloch sphere in Fig.~\ref{fig:Schematic}(b). 
As the system continues to squeeze, it reaches the state $\hat{\rho}_{\mathrm{peak}}$ which maximizes the QFI in time.
Here, the system is HL scaled with $\lambda_{\mathrm{max}} \sim 0.65 N^2$, and we discuss interesting properties of this state in Appendix~\ref{peakQFIappendix}.
In the large $N$ limit, we find that the maximum QFI asymptotes to $\lambda_{\mathrm{max}} \sim 0.64 N^2$, as shown in Fig.~\ref{fig:QFIplot}(b).
One can then rotate $\hat{\rho}_{\mathrm{peak}}$ to make a specific operator the optimal generator in order to exploit the largest amount of intraparticle entanglement for a specific sensing purpose~\cite{Reilly}. 
For example, in atomic clock systems, one would perform a $\pi / 2$ pulse about $(\hat{J}_x - \hat{J}_y) / \sqrt{2}$ in the rotating frame to make $\hat{J}_z$ the optimal generator.

While the PDD model can clearly reach a high QFI on an exponentially short timescale, the QCRB is not guaranteed to be achievable with experimentally accessible measurements. 
This is because the QFI implicitly optimizes over all measurement bases~\cite{Liu}.
Remarkably, the system saturates the QCRB with simple population measurements by performing a Bayesian reconstruction protocol. 
To demonstrate this, we first rotate the state by a $\pi / 2$ pulse with the optimal generator $\hat{\mathcal{G}}$ such that its anti-squeezed axis is parallel with the equator of the Bloch sphere. 
We encode the parameter $\Phi$ using $\hat{\mathcal{G}}$ and implement Bayes theorem $P(\Phi | m) = P(m | \Phi) P(\Phi) / P(m)$, where $P(\Phi|m)$ is a conditional probability for the outcome $m$ of a $\hat{J}_z$ measurement. 
We begin the process with a flat prior $P(m \lvert \Phi) = \mathrm{const.}$, which we then consistently update using the posterior distribution $P(\Phi \lvert m)$~\cite{Chih,Reilly3}. 

Figure~\ref{fig:QFIplot}(c) displays the sensitivity of the posterior distribution for the states $\hat{\rho}_{\mathrm{BW}}$ and $\hat{\rho}_{\mathrm{peak}}$ after the rotation to the Bloch sphere's equator.
We also show the SQL and HL as the upper and lower dotted lines. 
After $M$ measurements, the respective QCRBs are given by $1 / \sqrt{M \lambda_{\mathrm{max}} (t)}$, which we plot as dashed lines.
Remarkably, the sensitivity of $\hat{\rho}_{\mathrm{peak}}$ nearly reaches the HL.
For $\hat{\rho}_{\mathrm{peak}}$, the standard deviation $\sigma$ of the posterior distribution $P(\Phi | m)$ saturates this bound when $M \gtrsim 100$, showing that simple quadrature measurements are optimal for the generated states.
In contrast, we find that $\sigma$ of $P(\Phi|m)$ for $\hat{\rho}_{\mathrm{BW}}$ is a straight line when plotted in a $\log$-$\log$ plot as function of the measurement number. 
This highlights that every single measurement in the optimal measurement basis using the BW phase state has the smallest possible uncertainty in the parameter estimation of $\Phi$~\cite{Berry,Berry2}.
Remarkably, this means simple quadrature measurements for a state generated by the PDD model, $\hat{\rho}_{\mathrm{BW}}$, already saturates the QCRB after a single measurement which is an important property when developing practical, real-time inertial sensors.
We can calculate the decibel gain over the SQL, $G = 10 \log_{10} (\sqrt{\lambda_{\mathrm{max}} / N})$, and obtain $G = 5.7 \, \mathrm{dB}$ and $G = 9.1 \, \mathrm{dB}$ of squeezing for $\hat{\rho}_{\mathrm{BW}}$ and $\hat{\rho}_{\mathrm{peak}}$, respectively.
For $\hat{\rho}_{\mathrm{peak}}$ in the large $N$ limit, we expect the gain to scale as $G \approx 5 \log_{10}(N) - 1$.

To highlight the achievements of this squeezing protocol, we will now compare them to one-axis twisting (OAT). 
We show that, with respect to OAT, the periodically driving scheme achieves higher QFI and is faster. 
To demonstrate this, we now consider $\omega = 0$ in Eq.~\eqref{H_Dicke} and eliminate the fast-oscillating $\hat{J}_{\pm}^2$ terms via the RWA.
This gives the one-axis twisting (OAT) Hamiltonian~\cite{Kitagawa},
\begin{equation} \label{H_OAT}
    \hat{H}_{\mathrm{OAT}} \approx - \frac{\hbar \chi}{2} \hat{J}_z^2,
\end{equation}
as exploited in Refs.~\cite{Luo,ActuallyJarrodsPaper2}.
Here, we have used the relation $\hat{J}_+ \hat{J}_- = \hat{J}^2 - \hat{J}_z^2 + \hat{J}_z$ and ignored a constant energy shift of $N(N / 2 + 1) / 2$ from the $\hat{J}^2$ term since we remain in the collective subspace $\{ \ket{j = N/2, m}, \; - j \leq m \leq j \}$.

When the state begins in an eigenvector of $\hat{J}_x$, $\hat{\rho}_0 = [(\ket{\uparrow} + \ket{\downarrow})(\bra{\uparrow} + \bra{\downarrow}) / 2]^{\otimes N}$, the OAT Hamiltonian reaches $\lambda_{\mathrm{max}} = N (N + 1) / 2$ on a timescale of $t_{\mathrm{pl}} \sim 4 / (\sqrt{N} \abs{\chi})$~\cite{Pezze,Pezze2,Reilly}.
We show this initial behavior of the maximum QFI for OAT as a dashed black line in Fig.~\ref{fig:QFIplot}(a).
The QFI then remains at this value for a long plateau before eventually growing again at $t_{\mathrm{pl},f} \sim \pi / \abs{\chi} - 4 / (\sqrt{N} \abs{\chi})$~\cite{Pezze,Pezze2,Reilly}. 
For typical parameters, this is often too long of a timescale since decoherence will significantly reduce the squeezing performance.
We compare the typical timescales for HL scaling of PDD and OAT  in Fig.~\ref{fig:QFIplot}(d).
We find that the PDD model indeed scales on a much faster timescale, an observation which becomes more pronounced if one considers larger atom numbers.
Fitting the scaling of the PDD model, we find that the time that QFI is maximized is given by
\begin{equation} \label{t_peak}
    t_{\mathrm{peak}} \approx [\ln(N^2) + 4] / (N \abs{\chi}),
\end{equation}
which approximately matches the analysis of Ref.~\cite{Andre} with the Wineland squeezing parameter.
Therefore, the PDD model is a full order of magnitude faster than OAT when one scales up to $N = 10^4$ while reaching a higher QFI, as shown in Figs.~\ref{fig:QFIplot}(b) and~\ref{fig:QFIplot}(d).
Moreover, the states created by OAT do not, in general, saturate the QCRB using simple quadrature measurements when encoding the optimal parameter $\Phi$.
This shows that parametric driving is a very promising tool for fast spin squeezing.

\section{Example Experimental Realization \label{Sec:Exp}}

\subsection{Vertical Cavity Setup}
\begin{figure}
    \centerline{\includegraphics[width=\linewidth]{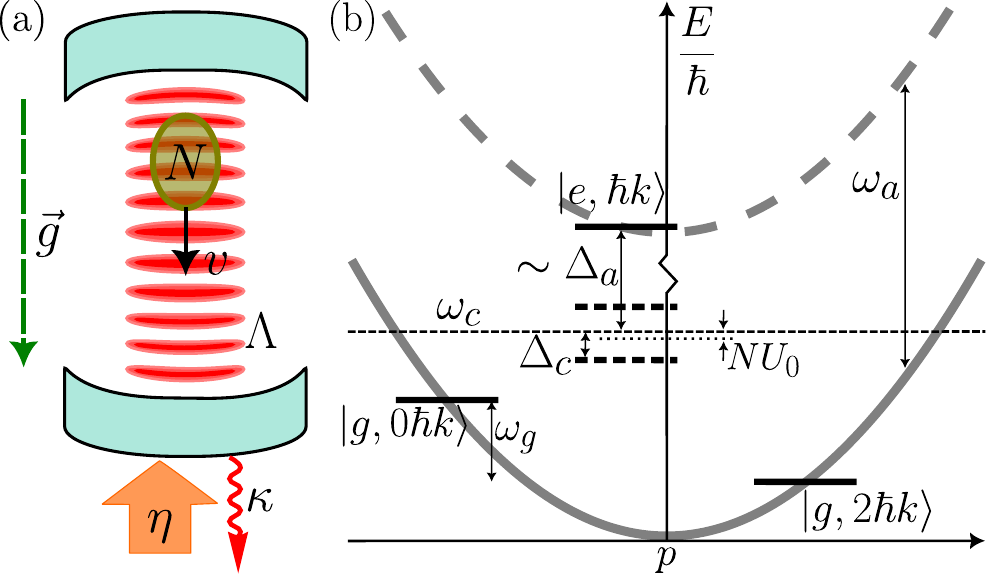}}
    \caption{(a) A schematic of the vertical cavity experiment of Refs.~\cite{Greve,Luo}.
    (b) Frequency spectrum diagram of a single atom in the vertical cavity setup.
    The states are labeled by their internal state and their initial momentum value, i.e., $\ket{i,p_0 - m g \tau} \rightarrow \ket{i,p_0}$.}
    \label{fig:VCschematic}
\end{figure}
Having established that the PDD model can outperform OAT on short timescales, we now turn to a prototypical experimental realization of this scheme.
For this, momentum squeezing in a recent vertical cavity (VC) experiment~\cite{Greve,Luo,ActuallyJarrodsPaper2,Zhang2} is considered, as shown schematically in Fig.~\ref{fig:VCschematic}(a).
A wave packet of $N$ ${}^{87} \mathrm{Rb}$ atoms falls through an optical mode of a VC under the influence of gravity $\vec{g} = -g \vec{\varepsilon}_z$ with unit vector $\vec{\varepsilon}_z$ along the vertical axis.
The atoms interact with a single mode of the VC, which has frequency $\omega_c$, via the $D_2$ cycling transition $\ket{g} \equiv \ket{F = 2, m_F = 2} \leftrightarrow \ket{e} \equiv \ket{F' = 3, m_{F'} = 3}$ at a single atom vacuum coupling rate $\Lambda$.
The internal states are separated by an optical frequency $\omega_a$ and $\ket{e}$ can decay back to $\ket{g}$ at a rate $\gamma$. 
It is assumed that the cavity decays at a rate $\kappa$ and initially begins in vacuum.

After the atoms have fallen for a time $\tau$, a coherent field is injected into the cavity which drives the mode with a time-dependent rate $\abs{\eta(t)}$ and frequency $\omega_p(t)$. 
The general idea is that the frequency $\omega_p(t)$ is modulated in order to achieve the parameteric driving necessary for the fast spin squeezing.
The modulation of the frequency $\omega_p (t) = \omega_p^{(0)} + \omega_p^{(1)} (t)$ is chosen to be around a frequency $\omega_p^{(0)}$ with a fixed detuning to the atoms $\Delta_a = \omega_a - \omega_p^{(0)}$ and the cavity $\Delta_c = \omega_c - \omega_p^{(0)}$. 
In a reference frame rotating with $\omega_p^{(0)}$, the dynamics of the atoms and cavity is described by a master equation for the density matrix $\hat{\rho}_{\mathrm{apc}}$ which takes the form
\begin{equation} \label{MasterEqAPC}
    \begin{aligned}
\pd{\hat{\rho}_{\mathrm{apc}}} = & - \frac{i}{\hbar} \left[ \hat{H}_{\mathrm{apc}}, \hat{\rho}_{\mathrm{apc}} \right] + \hat{\mathcal{D}} \left[ \sqrt{\kappa} \hat{a} \right] \hat{\rho}_{\mathrm{apc}} \\
&+ \sum_j \hat{\mathcal{D}} \left[ \sqrt{\gamma} \hat{\sigma}_j^- \right] \hat{\rho}_{\mathrm{apc}}.
    \end{aligned}
\end{equation}
The coherent dynamics is governed by the Hamiltonian
\begin{widetext}
\begin{equation} \label{Hapc}
    \hat{H}_{\mathrm{apc}} = \sum_j \left[ \frac{(\hat{p}_j - m g \tau)^2}{2 m} + \hbar \Lambda \cos(k \hat{x}_j) (\hat{a}^{\dagger} \hat{\sigma}_j^- + \hat{\sigma}_j^+ \hat{a}) + \hbar \Delta_a \hat{\sigma}_j^+ \hat{\sigma}_j^- \right] + \hbar \Delta_c \hat{a}^{\dagger} \hat{a} + \hbar \left[ \eta(t) \hat{a}^{\dagger} + \hat{a}\eta(t)^* \right].
\end{equation}
\end{widetext}
The first term in the Hamiltonian describes the kinetic energy with momentum operators $\hat{p}_j$ of the atoms with mass $m$ after falling for a time $\tau$ under acceleration $g$. 
The second term corresponds to the atom-cavity coupling, where $\cos(k x)$ is the standing-wave mode function of the cavity evaluated at the atomic position operators~$\hat{x}_j$ with wavenumber $k$. 
In addition, the creation and annihilation operators $\hat{a}^{\dagger}$ and $\hat{a}$ have been introduced for the cavity mode and the internal excitation $\hat{\sigma}_j^+ = \ket{e}_j \bra{g}_j$ and $\hat{\sigma}_j^- = \ket{g}_j \bra{e}_j$, respectively. 
The third term in Eq.~\eqref{Hapc} is the energy of the excited state in the frame rotating with $\omega_p^{(0)}$. 
The last two terms describe the energy of the photons and the driving of the cavity mode where modulations of frequency $\omega_p^{(1)} (t)$ and amplitude $\abs{\eta (t)}$ are encoded in the complex and time-dependent frequency $\eta (t)$.
This time-dependent $\eta(t)$ is crucial for the parameteric driving.
As well as the coherent effects, the master equation Eq.~\eqref{MasterEqAPC} also describes cavity photon losses and spontaneous emission as Markovian processes, and therefore these dissipative processes are modeled using the Lindblad superoperator 
\begin{equation} \label{LindbladDissipator}
    \hat{\mathcal{D}} [\hat{O}] \hat{\rho} = \hat{O} \hat{\rho} \hat{O}^{\dagger} - \frac{1}{2} [\hat{O}^{\dagger} \hat{O} \hat{\rho} + \hat{\rho} \hat{O}^{\dagger} \hat{O}].
\end{equation}
The master equation~\eqref{MasterEqAPC} is the starting point of the derivation which, after making several approximations discussed in the subsequent subsection, results in an effective Dicke model.

\subsection{Simplification of the Model}
While Eq.~\eqref{MasterEqAPC} represents a full quantum model of the setup shown in Fig.~\ref{fig:VCschematic}(a), an exact numerical simulation is infeasible except for the smallest atom numbers and low intracavity photon numbers. 
In order to develop an exact but efficient simulation that accurately models the dynamics of Eq.~\eqref{MasterEqAPC} for substantial atom numbers, we now simplify the model to a numerically tractable one by adiabatically eliminating different degrees of freedom of the system. 
This is possible when a clear separation of timescales exist, and one can tract the slow dynamics of one degree of freedom while ignoring the fast oscillating dynamics of another. 
To this end, we now discuss the regimes in which the excited state manifold and cavity field can be adiabatically eliminated, ending with a master equation that only depends on the external degree of freedom of the atoms.
We then simplify this further by truncating momentum space to two relevant states, ending in the final master equation Eq.~\eqref{MasterEq}.

\subsubsection{Adiabatic Elimination of Excited State Manifold}
The regime considered is that in which the detuning $|\Delta_a|$ is much larger than the spontaneous emission rate~$\gamma$ and any characteristic frequency determining the dynamics of the cavity and the atomic external degrees of freedom. 
In this regime, the atoms remain, to good approximation, in the electronic ground state and the dominant scattering process is coherent scattering of laser photons. 
In addition, it is assumed that the fixed atom-laser detuning is much larger than the dynamical variance of the frequency $|\Delta_a| \gg \omega_p^{(1)}$, which implies that the small modifications in the laser frequency have only a minor effect on the coherent scattering rates.
Using these approximations, based on the parameter regime of interest, an effective master equation is derived that governs the dynamics of the density matrix $\hat{\rho}_{\mathrm{pc}}$ of atomic external degrees of freedom and the cavity. 
This master equation is given by
\begin{equation}
    \pd{\hat{\rho}_{\mathrm{pc}}} = - \frac{i}{\hbar} \left[ \hat{H}_{\mathrm{pc}}, \hat{\rho} \right] +  \hat{\mathcal{D}} \left[ \sqrt{\kappa} \hat{a} \right] \hat{\rho}_{\mathrm{pc}},
\end{equation}
with the Hamiltonian~\cite{ActuallyJarrodsPaper2}
\begin{equation} \label{Hpc}
    \begin{aligned}
\hat{H}_{\mathrm{pc}} = & \sum_j \frac{(\hat{p}_j - m g \tau)^2}{2 m} + \hbar \left[ \eta(t) \hat{a}^{\dagger} + \mathrm{H.c.} \right] \\
&+ \hbar \Delta_c' \hat{a}^{\dagger} \hat{a} - \hbar U_0 \sum_j \cos(2 k \hat{x}_j) \hat{a}^{\dagger} \hat{a}.
    \end{aligned}
\end{equation}
The second line in Eq.~\eqref{Hpc} describes the modified frequency of cavity photons which is shifted due to the presence of the atoms. 
Here, $\Delta_c' = \Delta_c - N U_0$ is the dressed cavity detuning with the ac Stark shift 
\begin{equation}
    U_0 = \frac{1}{2} \frac{\Lambda^2 \Delta_a}{\Delta_a^2 + \gamma^2 / 4}.
\end{equation}

\subsubsection{Adiabatic Elimination of Cavity Field}
To simplify the model further, it is useful to note that the cavity field can be decomposed into two components.
First, the driving laser injects a field $\beta$ into the cavity, which comes from the $\Delta_c'$ and $\eta$ terms in Eq.~\eqref{Hpc}, as well as cavity decay.
Importantly, this injected field is present even when there are no atoms in the system.
On top of this, there is the field which is scattered off of the atoms in the two-photon process modeled by the final term in Eq.~\eqref{Hpc}.
Since $\beta (t)$ is, in principle, known for a certain time-dependent driving profile $\eta (t)$ and does not depend on the dynamics of the atoms, it can be eliminated from the model in order to focus on the scattered light field.  
This is formally done by applying the displacement transformation 
\begin{equation}
    \hat{D}_1 = \exp[\hat{a}^{\dagger} \beta(t) - \beta^* (t) \hat{a}],
\end{equation}
onto the density matrix $\tilde{\hat{\rho}}_{\mathrm{pc}} = \hat{D}_1^{\dagger} \hat{\rho}_{\mathrm{pc}} \hat{D}_1$. 
In this new displaced picture, we find
\begin{equation} \label{DisplacedMasterEq}
    \pd{\tilde{\hat{\rho}}_{\mathrm{pc}}} = - \frac{i}{\hbar} \left[ \tilde{\hat{H}}_{\mathrm{pc}}, \tilde{\hat{\rho}}_{\mathrm{pc}} \right] + \hat{\mathcal{D}} \left[ \sqrt{\kappa} \hat{a} \right] \tilde{\hat{\rho}}_{\mathrm{pc}},
\end{equation}
where the injected light field follows the driven-damped harmonic oscillator differential equation
\begin{equation} \label{BetaEtaDiffeq}
    \pd{\beta} = - i \left( \Delta_c' - \frac{i \kappa}{2} \right) \beta - i \eta.
\end{equation}
With a solution of Eq.~\eqref{BetaEtaDiffeq}, the following displaced Hamiltonian is obtained
\begin{widetext}
\begin{equation} \label{Hpcavnew}
    \tilde{\hat{H}}_{\mathrm{pc}} = \sum_j \left[ \frac{(\hat{p}_j - m g \tau)^2}{2 m} - \hbar U_0 \cos(2 k \hat{x}_j)(\hat{a}^{\dagger} \beta + \beta^* \hat{a}) - \hbar U_0 \abs{\beta}^2 \cos(2 k \hat{x}_j) \right] + \hbar \Delta_c' \left[ 1 - \epsilon \sum_j \cos(2 k \hat{x}_j) \right] \hat{a}^{\dagger} \hat{a},
\end{equation}
\end{widetext}
where $\epsilon = U_0 / \Delta_c'$.
Now in this displaced picture, there is no external driving of the cavity. 
Instead, the operator $\hat{a}$ represents the scattered field due to the presence of atoms which, in the original picture, needs to be added to the injected field $\beta$. 

In what follows, we work in the regime of large effective cavity detuning.
By assuming $\abs{\Delta_c'}$ is the largest frequency in the effective system, we are able to adiabatically eliminate the scattered cavity field. 
This requires that $\abs{\Delta_c'}$ is much larger than the Doppler-shift of the atoms and also that the modulation of the drive is slow compared to $1 / \abs{\Delta_c'}$. 
In this limit, we can derive an effective master equation for the density matrix describing the atomic external degrees of freedom $\hat{\rho}$, where we dropped the ``$\mathrm{p}$'' index for brevity.
Eliminating the bosonic field in Appendix~\ref{CavityElimAppendix} using the method of Ref.~\cite{Jager} results in the atom-only master equation 
\begin{equation} \label{MasterEqP}
    \pd{\hat{\rho}} \approx - \frac{i}{\hbar} [\hat{H}_{\mathrm{VC}}, \hat{\rho}] + \hat{\mathcal{D}} \left[ \sqrt{\Gamma_c(t)} \sum_j \cos(2 k \hat{x}_j) \right] \hat{\rho},
\end{equation}
where the assumption is made that $N \abs{\epsilon} / 2 \ll 1$.
Here, the Hamiltonian is given by 
\begin{equation} \label{Hp}
    \begin{aligned}
\hat{H}_{\mathrm{VC}} \approx & \sum_j \left[ \frac{(\hat{p}_j - m g \tau)^2}{2 m} - \hbar  U_0 \abs{\beta}^2 \cos(2 k \hat{x}_j) \right] \\
&- \hbar  \chi (t) \sum_{i,j} \cos(2 k \hat{x}_i) \cos(2 k \hat{x}_j),
    \end{aligned}
\end{equation}
and the nonlinear interaction rate is defined as
\begin{equation} \label{chiEq}
    \chi (t) = \frac{\Delta_c'U_0^2 \abs{\beta}^2}{(\Delta_c')^2 + \kappa^2 / 4},
\end{equation}
with the dissipation rate given by
\begin{equation} \label{GammaEq}
    \Gamma_c (t) = \frac{\kappa U_0^2 \abs{\beta}^2}{(\Delta_c')^2 + \kappa^2 / 4}.
\end{equation}
Furthermore, as discussed in Appendix~\ref{InjectedFieldAppendix}, the injected field can be approximated as 
\begin{equation}
    \beta (t) \approx - \frac{\eta (t)}{\Delta_c' - \frac{i \kappa}{2}},
\end{equation}
under suitable parameters.

Equation~\eqref{MasterEqP} solely and effectively describes the dynamics of the atoms. 
In this picture the cavity mediates long-range atom-atom interactions that are described by the third term in Eq.~\eqref{Hp}. 
This is a key ingredient of the approach: the cavity here is used to create a non-linearity in the atomic cloud. 
Moreover, the possibility to modulate $\eta$, and therefore $\beta$, in time allows the interaction strength in Eq.~\eqref{chiEq} to be driven. 
This combination of cavity-mediated interactions and in the case considered time-periodic parametric driving allows for the fast squeezing. 

Formally, the master equation~\eqref{MasterEqP} models a matter-wave system that can be simulated using the algebra of the group $\mathrm{SU} (n)$ when the atoms are taken to be permutationally symmetric. 
Here, $n$ corresponds to the number of relevant momentum states per atom after discretizing the momentum basis in units of $2 \hbar k$. 
Therefore, the scaling of the density matrix with respect to atomic number is reduced from exponential to polynomial, going approximately as $[N^{n - 1}/(n - 1)!]^2$~\cite{Mathur,Silva,Reilly4}.

\subsubsection{Reduction to Two Momentum States}
The specific experimental setup allows for futher simplification of the theoretical model since the momentum basis can be reduced to two relevant states in a certain parameter regime~\cite{Luo}.
First, the atoms are prepared in the momentum ground state $\ket{0 \hbar k}$.
By letting the atomic packet fall for a sufficient time $\tau$ before turning on the injected field, the momentum states $\ket{0 \hbar k}$ and $\ket{2 \hbar k}$ become nearly degenerate in the co-falling reference frame~\cite{ActuallyJarrodsPaper2}, with a difference in frequency of
\begin{equation} \label{omegaGdef}
    \omega_g = \frac{(2 \hbar k - m g \tau)^2 - (m g \tau)^2}{2 \hbar m} = 4 \omega_r - 2 k g \tau,
\end{equation}
with recoil frequency $\omega_r = \hbar k^2 / (2 m)$.
Therefore, under the conditions $12 k g \tau \gg \abs{U_0} \abs{\beta}^2$ and $16 \omega_r \gg N \abs{\chi}$, one can drive Bragg transitions between $\ket{0 \hbar k} \leftrightarrow \ket{2 \hbar k}$ while being energetically far from coupling to the $\ket{-2 \hbar k}$ and $\ket{4 \hbar k}$ states, truncating the momentum space to a collective two-level system.
The details of this process and the derivation of these conditions are presented in Appendix~\ref{TwoLevelApproxAppendix}.
Defining the collective momentum operators $\hat{J}_+ = \sum_j \op{2 \hbar k}{0 \hbar k}_j = \hat{J}_-^{\dagger}$, the two-level truncation reduces Eq.~\eqref{Hp} to
\begin{equation} \label{TruncatedH}
    \hat{H}_{\mathrm{VC}} \approx \hbar \omega_g \hat{J}_z - \hbar \chi (t) \hat{J}_x^2,
\end{equation}
where the $12 k g \tau \gg \abs{U_0} \abs{\beta}^2$ condition has been used to ignore the single atom momentum flip term in Eq.~\eqref{Hp}.
We present a table outlining the various approximations that we assume to be valid to derive Eq.~\eqref{TruncatedH} in Appendix~\ref{expParamsAppendix}, as well as relevant experimental parameters that satisfy these conditions which we use for our dynamical simulations in Fig.~\ref{fig:VCplot}.

\subsection{Realization of PDD Dynamics} \label{RealizationPDDdynamics}
In order to obtain the PDD Hamiltonian in Eq.~\eqref{H_Dicke}, we can now reverse engineer the driving profile $\eta(t)$.
For this, we require $\beta (t) = \beta_0 \sqrt{\cos(\omega t)}$ and so we set $\eta \propto \sqrt{\cos(\omega t)}$ which amounts to varying the amplitude and phase of the driving laser.
However, since $\chi \propto \abs{\beta}^2$, this does not yet have the needed harmonics to parameterically drive TACT in Eq.~\eqref{TruncatedH}.
Therefore, the cavity detuning is oscillated such that $\Delta_c' (t) = \Delta_c' (0) \sgn[\cos(\omega t)]$.
This promotes $\abs{\cos(\omega t)} \rightarrow \cos(\omega t)$ whereupon one sets $\omega = 2 \omega_g$. 
The oscillation of $\Delta_c'$ can be accomplished with a time-dependent pump frequency or with time-dependent laser powers when one adds a second pump laser with shifted frequency.
With this oscillation, Eq.~\eqref{TruncatedH} reduces to the PDD model of Eq.~\eqref{H_Dicke},
\begin{equation} \label{H_VC}
    \hat{H}_{\mathrm{VC}} = \hbar \omega_g \hat{J}_z - \hbar \chi_0 \cos(\omega t) \hat{J}_x^2,
\end{equation}
where $\chi_0 = U_0^2 \abs{\beta_0}^2 \Delta_c' (0) / ([\Delta_c' (0)]^2 + \kappa^2 / 4)$.
Then Eq.~\eqref{MasterEqP} can be rewritten to obtain the final master equation 
\begin{equation} \label{MasterEq}
    \pd{\hat{\rho}} = - \frac{i}{\hbar} \left[ \hat{H}_{\mathrm{VC}} (t), \hat{\rho} \right] + \hat{\mathcal{D}} \left[ \sqrt{\Gamma_0 \abs{\cos(\omega t)}} \hat{J}_x \right] \hat{\rho},
\end{equation}
with the collective decay rate amplitude from Eq.~\eqref{GammaEq},
\begin{equation}
    \Gamma_0 = \frac{\kappa U_0^2 \abs{\beta_0}^2}{(\Delta_c')^2 + \frac{\kappa^2}{4}}.
\end{equation}

This concludes the derivation of the master equation and adds physical meaning to the model parameters introduced in Eq.~\eqref{H_Dicke}. 
It also predicts the presence of dissipation that is automatically introduced by the intrinsic lifetime of cavity photons in such systems. 
In what follows, we will discuss how this dissipation effects the squeezing that was discussed in Sec.~\ref{Sec:Dicke}.

\section{Squeezing in The Presence of Dissipation \label{Sec:Diss}}

We will now discuss how the idealized squeezing that we described in Sec.~\ref{Sec:Dicke} is modified by realistic dissipation included in the description in Eq.~\eqref{MasterEq}, derived in detail in Sec.~\ref{Sec:Exp}. 
The presence of this quantum noise results in decoherence in the system and leads to mixed states, in contrast to the pure states that have been described in Sec.~\ref{Sec:Dicke}. 
In our protocol, it is in general beneficial to reduce the amount of dissipation which can be achieved by reducing the ratio of $\kappa/\Delta_c'$. 
Nevertheless, we will include finite dissipation in order to study the detrimental effect of decoherence on spin squeezing. 
The mathematical framework now becomes somewhat more involved as we have to work with mixed states. 
However, we can use the framework of optimal generators and QFIM described by Eq.~\eqref{OptimalGenerator} and Eq.~\eqref{QFIM} since we have already presented the QFIM in its mixed state form (as opposed to the covariant matrix form for pure states~\cite{Liu}). 

\begin{figure}
    \centerline{\includegraphics[width=0.8\linewidth]{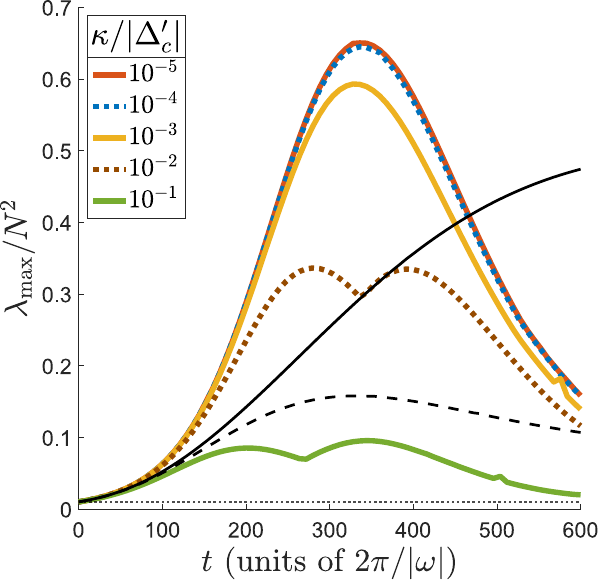}}
    \caption{The largest eigenvalue of the QFIM for the VC setup with different decay rates.
    The simulations evolve under Eq.~\eqref{MasterEq} with the full Hamiltonian of Eq.~\eqref{TruncatedH}.
    We choose the parameters $N = 100$, $\abs{\Delta_c} = 10.5 \abs{\omega_g}$, $\abs{U_0} = 5.1 \abs{\omega_g} \cross 10^{-3}$, and $\abs{\beta_0} = 6.8$ when $\kappa / \abs{\Delta_c'} \ll 1$, such that $\abs{\Delta_c'} = 9.9 \abs{\omega_g}$.
    We display the experimental parameters that lead to these rates in Appendix~\ref{expParamsAppendix}.
    The colored lines represent the PDD model ($\omega = 2 \omega_g$).
    Meanwhile, the black lines are the results of OAT ($\omega = 0$) for $\kappa / \abs{\Delta_c'} = 10^{-5}$ (solid black line) and $\kappa / \abs{\Delta_c'} = 10^{-2}$ (dashed black line).}
    \label{fig:VCplot}
\end{figure}

In Fig.~\ref{fig:VCplot}, the results for the maximum QFI, given by Eq.~\eqref{OptimalGenerator}, are displayed for a density matrix evolved under Eq.~\eqref{MasterEq} with different dissipation rates.
For comparison, results for OAT ($\omega = 0$) are also displayed with $\kappa / \abs{\Delta_c'} = 10^{-5}$ (solid black line) and $\kappa / \abs{\Delta_c'} = 10^{-2}$ (dashed black line). 
Notably, it is found that even with a three orders of magnitude larger dissipation rate, the PDD model ($\omega = 2 \omega_g$) outperforms OAT on short timescales, which can be seen by comparing the dotted brown line to the solid black line.
A potential explanation for the increased robustness of this realization of the PDD model is as follows. 
When one drives the cavity with a $\beta (t) \propto \sqrt{\cos(\omega t)}$ profile, the absolute values in Eqs.~\eqref{chiEq} and~\eqref{GammaEq} changes the harmonics to $\abs{\cos(\omega t)}$.
To recover the PDD model with parametric resonance at $\omega = 2 \omega_g$, we therefore had to oscillate $\Delta_c' (t)$ to promote $\chi \propto \abs{\cos (\omega t)} \rightarrow \cos(\omega t)$ as outlined in Sec.~\ref{RealizationPDDdynamics}.
However, the dissipation rate in Eq.~\eqref{GammaEq} does not have this factor of $\Delta_c'$, and therefore remains with the original harmonics of $\abs{\cos(\omega t)}$, as shown explicitly in Eq.~\eqref{MasterEq}.
This means that the $\Gamma_c (t) \hat{J}_x^2$ terms in the Lindblad dissipator Eq.~\eqref{LindbladDissipator} are not driven at the parametric resonance, and so do not have the exponentially speed up that the squeezing is awarded.
Further investigation into this advantageous property of the PDD model is left to future work. 

Some of the generated state's properties discussed in Appendix~\ref{peakQFIappendix} [see Fig.~\ref{fig:RhoPeak}] may explain another interesting feature of the dissipative dynamics that allows the system to remain highly sensitive for longer periods of time, namely the double peak structure of the PDD model in Fig.~\ref{fig:VCplot} when $\kappa / \abs{\Delta_c'} \gtrsim 10^{-2}$.
Here, the first peak corresponds to the initial squeezing with $\hat{\mathcal{G}} = (\hat{J}_x + \hat{J}_y) / \sqrt{2}$, but now with a lower QFI that reaches its maximum value more quickly due to decoherence. 
The optimal generator then switches to $\hat{J}_z$ for the second peak as the QFI with respect to $\hat{J}_z$ rotations falls off less quickly when increasing $\kappa$.
This generator corresponds to the second largest eigenvalue of the QFIM in Fig.~\ref{fig:QFIplot}(a) (blue dashed line), and so the maximum of this QFI is also decreasing (from the original value of $\lambda_2 \approx 0.4 N^2$) and occurring sooner with dissipation, but at a slower rate than the first peak.
This is most likely due to the partial-ring structure of the generated states [see Fig.~\ref{fig:RhoPeak}(a)] being more robust to decoherence than the state's fringes [see Fig.~\ref{fig:RhoPeak}(b)] as a large amount of the population is near the eigenstates of the jump operator in Eq.~\eqref{MasterEq}, and these population packets are more metrologically useful for rotations about $\hat{J}_z$ than about a point on the Bloch sphere's equator, i.e.~$\hat{\mathcal{G}}$.

To put the results into an experimental context, we adopt the setup of Refs.~\cite{Luo,ActuallyJarrodsPaper2} in which the atoms are allowed to fall for $\tau = 20 \, \mathrm{ ms}$ before the pump is turned on. This corresponds to $\abs{\omega_g} \sim 2 \pi \cross 0.5 \, \mathrm{MHz}$ such that $\abs{\omega} \sim 2 \pi \cross 1 \, \mathrm{MHz}$.
Therefore, Fig.~\ref{fig:VCplot}(b) shows an appreciable advantage of the PDD model compared to OAT after $O(100 \, \mu s)$.
Furthermore, using the parameters of Fig.~\ref{fig:VCplot}(b) with small dissipation rates $\kappa / \abs{\Delta_c'} \ll 1$, it is found that $N \chi_0 \approx 0.012 \abs{\omega_g}$ and so Eq.~\eqref{t_peak} gives $t_{\mathrm{peak}} \sim 355 \, \mu \mathrm{s}$ while the OAT plateau time is $t_{\mathrm{pl}} \sim 1.1 \, \mathrm{ms}$.
On the timescale of $t_{\mathrm{peak}}$, an effective dephasing effect occurs from the increased energy difference between the momentum states as time progresses, which is accounted for in Ref.~\cite{Luo} by a spin echo sequence~\cite{Steck}.
Furthermore, this dephasing is a single-particle effect and so increasing $N$ can grow the collective squeezing rate without increasing the effective dephasing rate.
It is confirmed that the QCRB is saturated from the simple quadrature measurements considered in Sec.~\ref{Sec:Dicke}, which can be experimentally accessed via a Mach–Zehnder interferometry sequence~\cite{Greve}.
For the case of $\kappa \approx \abs{\Delta_c'} / 87$, which corresponds to the cavity decay of Ref.~\cite{Luo}, we find that the PDD model reaches a maximum of $G = 7.5 \, \mathrm{dB}$ during the initial squeezing.

\section{Conclusion and outlook \label{Sec:Conclusion}}
Similar to parametric driving of nonlinear optical interactions to create non-classical states of light~\cite{Kippenberg}, in this work, we propose an analogous procedure to create non-classical states of matter through parametric driving.
While we have focused on long-range interparticle interactions mediated through a dispersive cavity mode, our periodic driving methodology should be more broadly applicable to any system with controllable nonlinearities, such as trapped ions with phonon-mediated interactions~\cite{Sterk,Linnet,Wilson,Bohnet,Kahan}, Bose-Einstein condensates with short- and long-range interactions~\cite{Pyrkov,Kroeze,Mivehvar}, and solid state materials with spin-spin interactions~\cite{Zhou,Xie,Lee}.
Our periodic driving scheme is distinct from previous modulation proposals~\cite{Huang,Wu} as it is implemented by simple parameter modulation of classical driving fields, thereby allowing direct modulation of nonlinear Hamiltonian terms.
Unlike previous works on bosonic-mediated quantum amplification~\cite{Lu,Zeytinoglu,Burd}, the protocol presented here does not require squeezed bosonic modes and instead amplifies nonlinearities in the underlying matter to create non-classical, squeezed states.
We have demonstrated that our proposed method can potentially be implemented in a current, state-of-the-art VC experiment~\cite{Greve,Luo}, which would be the first experimental realization of TACT.
The system achieves HL scaling in reasonable timescales and has a simple optimal measurement basis, and therefore is a promising platform to create matterwave sensors with a true quantum advantage.
Furthermore, it has been shown~\cite{Kajtoch,Yukawa} that TACT creates the Berry-Wiseman phase state, as well as high overlap with other theoretically studied states~\cite{Yurke,Holland,Stockton,Combes,Huang2}.
Therefore, our proposal offers a promising platform to study previous theoretical work in quantum optics~\cite{Combes} and quantum information science~\cite{Berry,Berry2} in a controllable experimental spin system.

\begin{acknowledgments}
We thank Peter Zoller, Susanne Yelin, Catie LeDesma, and Kendall Mehling for useful discussions.
J.T.R., J.D.W, and M.J.H. acknowledge support from NSF PHY 2317149; NSF OMA 2016244; NSF PHY Grant No. 2207963; and NSF 2231377.
S.B.J. and S.E. acknowledge support from the Deutsche Forschungsgemeinschaft (DFG): Projects A4 and A5 in SFB/Transregio 185: “OSCAR”.
\end{acknowledgments}

\appendix

\section{States Created by the Periodically Driven Dicke Model}
In this section, we comment on some of the properties of two states that the periodically driven Dicke (PDD) model creates. 
We focus on the states examined in Fig.~\ref{fig:QFIplot}, namely, the Berry-Wiseman (BW) phase state $\hat{\rho}_{\mathrm{BW}}$ and the state with the peak QFI $\hat{\rho}_{\mathrm{peak}}$.

\subsection{Berry-Wiseman Phase State} \label{BWappendix}
We begin by discussing the BW phase state, whose Q-function is show in Fig~\ref{fig:Schematic}(b).
The Holevo variance for an ensemble of pseudospin-$1 / 2$ particles is defined as~\cite{Holevo,Berry}
\begin{equation} \label{HolevoVar}
    V (\varphi)_\psi \equiv |\langle e^{-i \varphi} \rangle_\psi|^2 - 1, 
\end{equation}
where
\begin{equation}
    \begin{aligned}
\langle e^{-i \varphi} \rangle_\psi &\equiv \int_0^{2 \pi} P_\psi(\varphi) e^{-i \varphi} d\varphi, \\
P_\psi(\varphi) &\equiv \bra{\psi} e^{-i \varphi \hat{J}_z} \ket{\psi}.
    \end{aligned}
\end{equation}
This variance is useful because states with complete phase uncertainty (e.g., any $\ket{\psi} = \ket{j = N / 2,m}$) will have infinite Holevo variance, whereas the typical phase variance~\cite{Berry}, $\Delta \varphi^2 = \langle\varphi^2\rangle_\psi - \langle\varphi\rangle_\psi^2$, has a maximum uncertainty of $\Delta\varphi = 2 \pi$.
It has been shown~\cite{Berry,Combes} that the state which minimizes the Holveo variance is the BW phase state,
\begin{equation} \label{BWstate}
   \ket{\psi_{\mathrm{BW}}} = \frac{1}{\sqrt{\frac{N}{2} + 1}} \sum_{m = -\frac{N}{2}}^{\frac{N}{2}} \sin \left[ \frac{\pi (\frac{N}{2} + m + 1)}{N + 2} \right] \ket{\frac{N}{2},m},
\end{equation}
such that $\hat{\rho}_{\mathrm{BW}} = \op{\psi_{\mathrm{BW}}}{\psi_{\mathrm{BW}}}$.
This state has $V(\varphi)_{\mathrm{BW}} = \pi^2 / N^2$ and is notably an eigenstate of the Susskind cosine operator~\cite{Susskind},
\begin{equation}
    \widehat{\cos} (\varphi) \equiv \frac{1}{2} \sum_{m = - N / 2}^{N / 2} \left( \op{\frac{N}{2},m+1}{\frac{N}{2},m} + \mathrm{H.c.} \right).
\end{equation}

The BW phase state is of particular interest for phase estimation because its dynamic range is a full $2 \pi$, meaning $\bra{\psi_{\mathrm{BW}}} e^{-i \varphi \hat{J}_z} \ket{\psi_{\mathrm{BW}}} = 1$ only if $\varphi = n 2 \pi$ for integer $n$.
Simultaneously, it has a quantum Fisher information (QFI) reaching Heisenberg limit (HL) scaling at
\begin{equation}
    \mathcal{F}_{\mathrm{BW}} \approx \left( \frac{1}{3} - \frac{2}{\pi^2} \right) N^2 \approx 0.13 N^2.
\end{equation}
These conditions guarantee that, with no a priori knowledge of $\varphi$, the BW phase state is the optimal state to gain information in a single measurement~\cite{Berry2}, making it a useful state for a multitude of sensing applications.
For example, creating a BW phase state in matterwave interferometry would guarantee that each measurement gives the highest resolution estimation of an acceleration, which would be a powerful tool for time-varying gravitational fields such as those that an orbiting satellite experiences.

\subsection{State with Peak QFI} \label{peakQFIappendix}
\begin{figure}
    \centerline{\includegraphics[width=\linewidth]{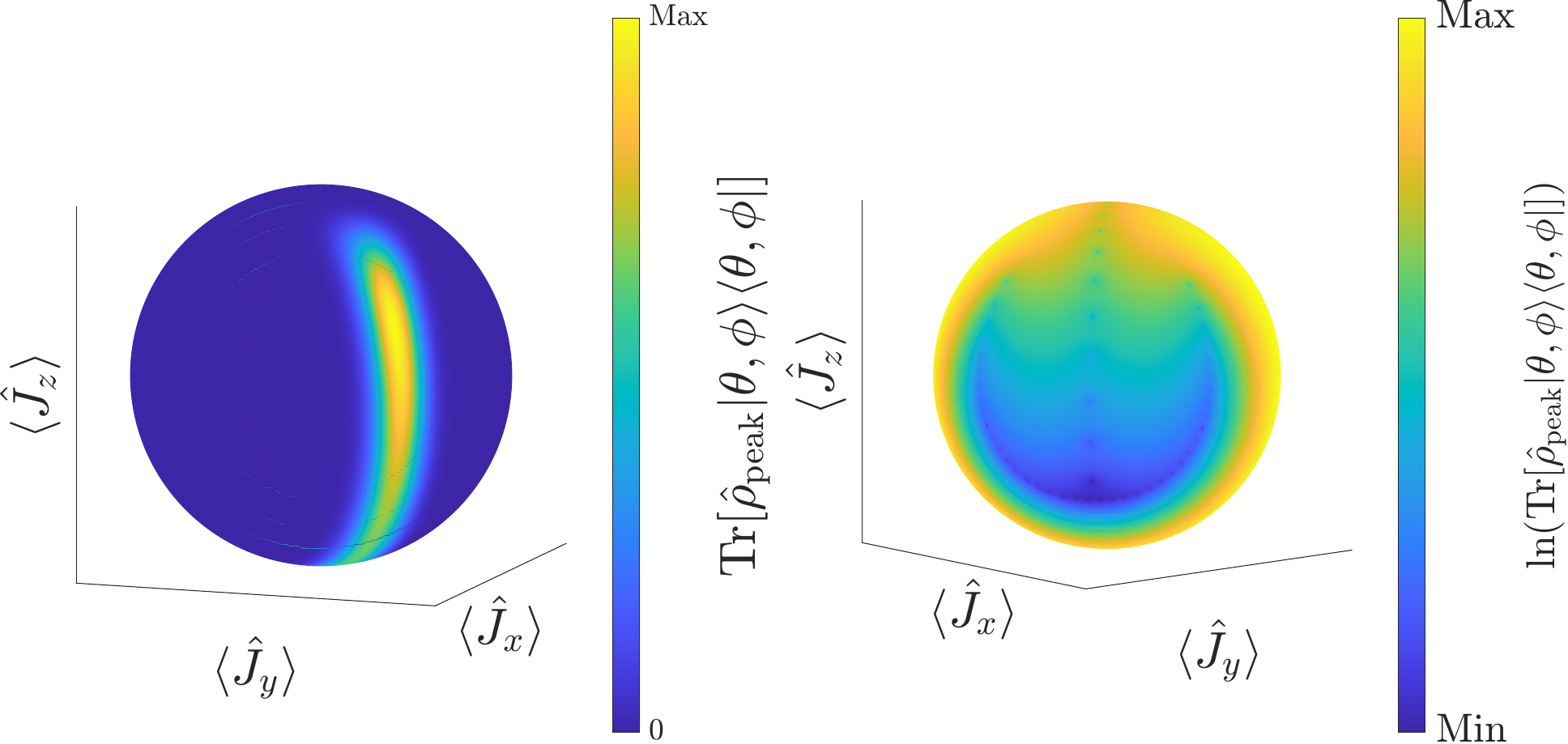}}
    \caption{The state with the maximum QFI $\hat{\rho}_{\mathrm{peak}}$ for $N = 100$. 
    (a) The Q-function calculated by finding the overlap with the coherent spin state $\ket{\theta,\phi}$ at every point on the Bloch sphere.
    (b) The log of the Q-function to make the interference fringes more pronounced.}
    \label{fig:RhoPeak}
\end{figure}
We now discuss the state with the maximum QFI during the initial squeezing under the PDD model, $\hat{\rho}_{\mathrm{peak}}$.
We display the Q-function of this state in Fig.~\ref{fig:RhoPeak}(a) which shows that $\hat{\rho}_{\mathrm{peak}}$ has properties of a partial ring state~\cite{Reilly2}.
One would expect this structure to be highly sensitive to rotations about $\hat{J}_z$ and a point on the Bloch sphere's equator in the direction of the anti-squeezed axis.
This explains why the two largest eigenvalues of the QFIM in Fig.~\ref{fig:QFIplot}(a) correspond to $\hat{\mathcal{G}} = (\hat{J}_x + \hat{J}_y) / \sqrt{2}$ and $\hat{\mathcal{G}}_2 = \hat{J}_z$ in the rotating frame.
Moreover, by taking a log of the Q-function, which we show in Fig.~\ref{fig:RhoPeak}(b), one can see interference fringes form around part of a longitude line of the Bloch sphere.
This is reminiscent of the interference fringes that are present in the $N00N$ state~\cite{Combes,Pezze} and may explain why the state is more sensitive to rotations about $\hat{\mathcal{G}}$ than $\hat{J}_z$.

As the squeezing continues past $\hat{\rho}_{\mathrm{peak}}$ under the PDD model with small dissipation, the large population packets begin to converge towards each other at the north pole.
However, the state's QFI remains larger than the SQL as interference fringes remain present with a small amount of population still in a partial ring.
The state reaches a local minimum in QFI when the large population packets meet at the north pole, but then the QFI climbs back to $\lambda_{\mathrm{max}} > N^2 / 2$ as a ring-like structure reemerges. 
This ring-like state has $\hat{\mathcal{G}} = \hat{J}_z$.

\begin{widetext}
\section{Adiabatic Elimination of the Cavity Field} \label{CavityElimAppendix}
We begin with Eq.~\eqref{DisplacedMasterEq} with the Hamiltonian from Eq.~\eqref{Hpcavnew}.
To eliminate the field, we assume that the scattered field is, to a good approximation, in vacuum.
We can then displace the field by 
\begin{equation}
    \hat{D}_2 = \exp[\hat{a}^{\dagger} \hat{\alpha} - \hat{\alpha}^{\dagger} \hat{a}],
\end{equation}
such that the equation of motion for $\hat{\rho}$ is given by~\cite{Jager}
\begin{equation} \label{MasterEq_full}
    \pd{\hat{\rho}} = - \frac{i}{\hbar} [\hat{H}_{\mathrm{VC}}, \hat{\rho}] + \hat{\mathcal{D}} \left[ \sqrt{\kappa} \hat{\alpha} \right] \hat{\rho},
\end{equation}
with the Hamiltonian
\begin{equation} \label{Hp_full}
    \hat{H}_{\mathrm{VC}} = \sum_j \left[ \frac{(\hat{p}_j - m g \tau)^2}{2 m} - \hbar  U_0 \abs{\beta}^2 \cos(2 k \hat{x}_j) \right] - \frac{\hbar U_0}{2} \left[ \beta \hat{\alpha}^{\dagger} \sum_j \cos(2 k \hat{x}_j) + \mathrm{H.c.} \right].
\end{equation}
We then solve for the effective field operator 
\begin{equation} \label{AlphaDiffEq}
    \pd{\hat{\alpha}} = - i \left[ \frac{(\hat{p}_j - m g \tau)^2}{2 \hbar m}, \hat{\alpha} \right] - i \left[ \Delta_c' \left( 1 - \epsilon \sum_j \cos(2 k \hat{x}_j) \right) - \frac{i \kappa}{2} \right] \hat{\alpha} + i U_0 \beta \sum_j \cos(2 k \hat{x}_j).
\end{equation}
Here, we have assumed that $U_0 \abs{\beta}^2$ is much smaller than any momentum energy gaps (see Section~\ref{TwoLevelApproxAppendix} for the relevant gaps) such that it can be dropped from the commutator in Eq.~\eqref{AlphaDiffEq}.

We are considering parameters such that $N \abs{\epsilon} / 2 \ll 1$ so that we can drop the non-linearity $\propto \epsilon$ in Eq.~\eqref{AlphaDiffEq}. 
By further making the ansatz $\hat{\alpha} (t) = a_+ (t) \sum_j \exp[2 i k \hat{x}_j] + a_- (t) \sum_j \exp[- 2 i k \hat{x}_j]$, we can find equations of motion for the coefficients $a_{\pm}$.
In the parameter regime $\abs{\Delta_c' - i \kappa / 2} \gg \omega$, where $\omega$ is the characteristic modulation frequency of $\beta$ [see Eqs.~\eqref{omega} and \eqref{omegaGdef}], we can integrate the differential equations for $a_{\pm}$. 
Using the obtained results in $\hat{\alpha} (t)$ leads to the effective field operator 
\begin{equation}
    \hat{\alpha} (t) \approx \frac{U_0 \beta}{2} \sum_j \left[ \frac{1}{\Delta_c' + \Delta p_{+ 2} - \frac{i \kappa}{2}} e^{2 i k \hat{x}_j} + \frac{1}{\Delta_c' + \Delta p_{- 2} - \frac{i \kappa}{2}} e^{- 2 i k \hat{x}_j} \right],
\end{equation}
where $\Delta p_{\pm 2} = (\hat{p} \pm 2 \hbar k - m g \tau)^2 / (2 \hbar m) - (\hat{p} - m g \tau)^2 / (2 \hbar m)$.

We now assume that we are restricted to low energy motional states, which we will formally justify in Sec.~\ref{TwoLevelApproxAppendix}.
For these states, we can set $\Delta_c' \pm \Delta p_{\pm 2} \approx \Delta_c'$ such that the effective field operator becomes
\begin{equation} \label{AlphaSimplified}
    \hat{\alpha} (t) \approx \frac{U_0 \beta}{\Delta_c' - \frac{i \kappa}{2}} \sum_j \cos(2 k \hat{x}_j).
\end{equation}
This is valid if $\Delta_c' \gg \Delta p_{\pm 2}$ and, for the situation considered here, amounts to $\Delta_c' \pm \omega_g \approx \Delta_c'$  where $\omega_g$ is given by Eq.~\eqref{omegaGdef} in Sec.~\ref{TwoLevelApproxAppendix}.
Using  Eq.~\eqref{AlphaSimplified} in Eqs.~\eqref{MasterEq_full} and~\eqref{Hp_full}, we find
\begin{equation}
    \pd{\hat{\rho}} \approx - \frac{i}{\hbar} [\hat{H}_{\mathrm{VC}}, \hat{\rho}] + \hat{\mathcal{D}} \left[ \sqrt{\frac{\kappa U_0^2 \abs{\beta}^2}{(\Delta_c')^2 + \kappa^2 / 4}} \sum_j \cos(2 k \hat{x}_j) \right] \hat{\rho},
\end{equation}
and the Hamiltonian
\begin{equation}
    \hat{H}_{\mathrm{VC}} \approx \sum_j \left[ \frac{(\hat{p}_j - m g \tau)^2}{2 m} - \hbar  U_0 \abs{\beta}^2 \cos(2 k \hat{x}_j) \right] - \frac{\hbar \Delta_c'U_0^2 \abs{\beta}^2}{(\Delta_c')^2 + \kappa^2 / 4} \sum_{i,j} \cos(2 k \hat{x}_i) \cos(2 k \hat{x}_j).
\end{equation}

\section{Reduction to Two Momentum States} \label{TwoLevelApproxAppendix}
In our protocol, the atoms are initialized with momentum $p = 0$, which means they have the kinetic energy $N m g^2 \tau^2 / 2$ after gravitational acceleration. 
The idea of the periodic driving with $\eta$ is now to engineer an injected light field $\beta$ which drives a pair creation process by flipping two momentum state to $p = 2 \hbar k$. 
This requires that we must drive with a frequency 
\begin{equation} \label{omega}
    \omega = 2 \omega_g,
\end{equation}
where $\omega_g$ denotes the energy to excite a single atom from $p = 0 \hbar k$ to the momentum state $p = 2 \hbar k$,
\begin{equation} 
    \omega_g = \frac{(2 \hbar k - m g \tau)^2 - (m g \tau)^2}{2 \hbar m} = 4 \omega_r - 2 k g \tau.
\end{equation}
Here, we have introduced the recoil frequency $\omega_r = \hbar k^2 / (2 m)$.
Thus, an appropriate driving profile would realize $\chi (t) \propto \cos(\omega t)$.
Using Eq.~\eqref{chiEq}, this can be realized with a driving resulting in $\abs{\beta(t)}^2 \propto \abs{\cos(\omega t)}$ and $\Delta_c' \propto \sgn[\cos(\omega t)]$, as explained in Section~\ref{RealizationPDDdynamics}. 
The latter corresponds to switching the driving frequency of the laser with respect to the cavity from red to blue detuned and back periodically in time.

We now want to restrict the dynamics of the atomic motional states to the momentum states $\ket{p = 0}$ and $\ket{p = 2 \hbar k}$. 
This requires that we do not excite other momentum states, which can be justified using time-dependent perturbation theory. 
The two most relevant momentum flips occur due to (a) the single-particle term proportional to $\cos(2k\hat{x}_j)$ in Eq.~\eqref{Hp} which induces the momentum flip of a single atom $p = \pm 2 \hbar k$, and (b) the two-particle term proportional to $\cos(2 k \hat{x}_i) \cos(2 k \hat{x}_j)$ in Eq.~\eqref{Hp} which can also amplify a pair with $p_1 = \pm 2 \hbar k$ and $p_2 = - 2 \hbar k$. 
We examine the requirements to avoid these two processes individually:

(a) The frequency gap for a single flip into the state $p = \pm 2 \hbar k$ is $\Delta \omega^{(1)}_{\pm}$. 
It can be calculated as
\begin{equation}
    \Delta \omega^{(1)}_{\pm} = \frac{(\pm 2 \hbar k - m g \tau)^2 - (m g \tau)^2}{2 \hbar m} = 4 \omega_r \mp 2 k g \tau.   
\end{equation}
The driving field $\abs{\beta}^2 \propto \abs{\cos(\omega t)}$ has frequency components that are multiples of $2 \omega = 4 \omega_g$. 
To neglect single momentum flips, we therefore require~\cite{Steck}
\begin{equation} \label{SingleFlipElim}
    \left| \frac{U_0 \abs{\beta}^2/2}{\abs{4\omega_g} - \abs{4 \omega_r \mp 2 k g \tau}} \right| \ll 1.  
\end{equation}
For large $k g \tau \gg 2 \omega_r$, this is true when $\abs{U_0} \abs{\beta}^2 \ll 12 k g \tau$.

(b) We now determine the frequency gap $\Delta \omega^{(2)}_{\pm}$ for the unwanted pair creation processes corresponding to creating $p_1 = \pm 2 \hbar k$ and $p_2 = - 2 \hbar k$. 
The frequency gap is given by
\begin{equation}
    \begin{aligned}
\Delta \omega^{(2)}_{+} &= \frac{(2 \hbar k - m g \tau)^2 - (m g \tau)^2}{2 \hbar m} + \frac{(- 2 \hbar k - m g \tau)^2 - (m g \tau)^2}{2 \hbar m} = 8 \omega_r, \\
\Delta \omega^{(2)}_{-} &= 2 \frac{(- 2 \hbar k - m g \tau)^2 - (m g \tau)^2}{2 \hbar m} = 8 \omega_r + 4 k g \tau.
    \end{aligned}
\end{equation}
Since we assume $\chi (t) \propto \cos(\omega t)$ with $\omega = 2 \omega_g$, these pair creation processes can be neglected if
\begin{equation} \label{DoubleFlipElim}
    \left| \frac{N \chi}{\abs{2 \omega_g} - \abs{\Delta \omega^{(2)}_{+}}} \right| \ll 1.
\end{equation}
Again assuming $k g \tau \gg 2 \omega_r$, this approximation is valid if $N \chi \ll 16 \omega_r$. 
In this calculation, we have included a factor of $N$ because of the collective enhancement. 

In the parameter regime where we can reduce the dynamics to atoms with momenta $p = 0$ and $p = 2 \hbar k$, we can identify the momentum raising operator as an effective collective spin raising operator
\begin{equation}
    \sum_j \exp[2 i k \hat{x}_j] \rightarrow \hat{J}_+ = \sum_j \ket{2 \hbar k}_j \bra{0 \hbar k}_j.
\end{equation}
We also define $\hat{J}_- = \hat{J}_{+}^{\dagger}$ as well as the $\mathrm{SU} (2)$ basis operators $\hat{J}_x = (\hat{J}^+ + \hat{J}^-) / 2$, $\hat{J}_y = i (\hat{J}_- - \hat{J}_+) / 2$, and $\hat{J}_z = [\hat{J}_+, \hat{J}_-] / 2$, where we note $\sum_j \cos( 2 k \hat{x}_j) \rightarrow \hat{J}_x$. 
With these definitions, we can rewrite the Hamiltonian in Eq.~\eqref{Hp} as the periodically driven Dicke (PDD) model
\begin{equation} \label{H_VC_supp}
    \begin{aligned} 
\hat{H}_{\mathrm{VC}} &= \hbar \omega_g \hat{J}_z - \hbar \chi (t) \hat{J}_x^2 \\
&= \hbar \omega_g \hat{J}_z - \hbar \chi_0 \cos(t) \hat{J}_x^2,
    \end{aligned}
\end{equation}
with $\chi_0 = U_0^2 \abs{\beta_0}^2 \Delta_c' (0) / ([\Delta_c' (0)]^2 + \kappa^2 / 4)$.
We also find a dissipative term with jump operator
\begin{equation}
    \hat{L} = \sqrt{\Gamma_c(t)} \hat{J}_x,
\end{equation}
with $\Gamma_c (t) \propto \abs{\cos(\omega t)}$.
\end{widetext}

\section{Profile of the Injected Field} \label{InjectedFieldAppendix}
We now comment on the driving profile of the injected field into the VC setup that reproduces the behavior of the periodically driven Dicke model.
We begin with the relationship between the injected field and standing field, Eq.~\eqref{BetaEtaDiffeq}.
Formally integrating and making a coarse-graining approximation, we find
\begin{equation} \label{BetaInt}
    \begin{aligned} 
\beta (t) &= e^{-i ( \Delta_c' - \frac{i \kappa}{2}) t} \beta(0) - i \int_0^t ds e^{-i ( \Delta_c' - \frac{i \kappa}{2}) s} \eta (t - s) \\
&\approx - \frac{\eta (t)}{\Delta_c' - \frac{i \kappa}{2}},
    \end{aligned}
\end{equation}
where we have assumed that the temporal variation of $\eta$ is slow compared to the exponential kernel in the integral. 
Within this limit, we can now reverse engineer $\eta (t)$ by simply inverting Eq.~\eqref{BetaInt}.

In the case that the coarse-graining approximation used in Eq.~\eqref{BetaInt} breaks down, one can instead plug $\beta (t) = \beta_0 \sqrt{\cos(\omega t)}$ into Eq.~\eqref{BetaEtaDiffeq} with the result
\begin{equation} \label{etaLong}
    \eta(t) = - \frac{i \omega \beta_0}{2} \sqrt{\sin(\omega t) \tan(\omega t)} - \beta_0 \left( \Delta_c' - \frac{i \kappa}{2} \right) \sqrt{\cos(\omega t)}.
\end{equation}
While the second term in this equation is the adiabatic result of the driving profile, the first term exhibits divergences which originate from the non-analyticities of $\sqrt{\tan(\omega t)}$. 
The first term contributes a factor of $\sqrt{\omega / \Delta_c'}$ in the integral for $\beta (t)$, Eq.~\eqref{BetaInt}.
In an integral over $\beta (t)$, it will be suppressed by a factor of $(\omega / \Delta_c')^{3/2}$~\cite{Gradshteyn}, and so the second term in Eq.~\eqref{etaLong} will be the dominate contribution. 
However, for experimental considerations, it might be advantageous to use different functions with smoother profiles that do not have such harsh intensity and phase control demands. 
We expect that these profiles can have similar performances for squeezing, although they might lead to shifted parametric resonances for $\omega$~\cite{Metelmann} which can be derived using a Holstein-Primakoff approximation~\cite{Holstein,Byrnes} for early times. 
For practical applications, it is also of interest to optimize $\eta (t)$ in order to achieve the maximum squeezing in minimum time with given experimental constraints.
These considerations are left for future work.

\section{Experimental Parameters} \label{expParamsAppendix}
In this section, we present experimental parameters that lead to the values of $\omega_g$, $\beta_0$, $U_0$, and $\Delta_c'$ used in Fig.~\ref{fig:VCplot}(b).
We use $N = 100$ throughout this section.
We then begin with the single-atom coupling constant of the cavity used in Ref.~\cite{Luo}, $\Lambda = 2 \pi \cross 0.5 \, \mathrm{MHz}$.
For this section, we also use the cavity loss rate from Ref.~\cite{Luo}, $\kappa = 2 \pi \cross 56 \, \mathrm{kHz}$.
The cavity addresses the $D_2$ cycling transition of ${}^{87} \mathrm{Rb}$, which is a $\lambda = 780 \, \mathrm{nm}$ transition with a decay rate of $\gamma = 2 \pi \cross 6.066 \, \mathrm{MHz}$~\cite{Steck2}.
We assume the injected field leads to a cavity pump rate $\eta_0 = 2 \pi \cross 33 \, \mathrm{MHz}$ and is detuned from the atomic resonance by $\abs{\Delta_a} = 2 \pi \cross 50 \, \mathrm{MHz}$.
The cavity frequency is also far detuned from the atomic resonance, while being detuned from the pump's frequency by $\abs{\Delta_c} = 2 \pi \cross 5.1 \, \mathrm{MHz}$.
Since all frequencies are within $O(100 \, \mathrm{MHz})$ from one another, we approximate the wavenumbers $k$ to be constant such that the recoil frequency from all photons in the system is approximated as $\omega_r = 2 \pi \cross 3.77 \, \mathrm{kHz}$~\cite{Steck2}.

With all of these specified experimental parameters, we obtain $\abs{U_0} = 2 \pi \cross 2.5 \, \mathrm{kHz}$, $\abs{\Delta_c'} = 2 \pi \cross 4.85 \, \mathrm{MHz}$, and $\abs{\beta_0} = 6.8$.
Furthermore, a drop time of $\tau = 20 \, \mathrm{ms}$ leads to $k g \tau = 2 \pi \cross 0.25 \, \mathrm{MHz}$ such that $\omega_g = - 2 \pi \cross 0.488 \, \mathrm{MHz}$.
This satisfies $k g \tau \gg 2 \omega_r$, which was used in Eqs.~\eqref{SingleFlipElim} and~\eqref{DoubleFlipElim}, by a factor of $33$.
We thus have all of the needed quantities to simulate Eq.~\eqref{H_VC_supp}.
We can also calculate the perturbation $\abs{\epsilon} = 5.1 \cross 10^{-4}$, standing field $\abs{U_0} \abs{\beta_0}^2 = 2 \pi \cross 0.115 \, \mathrm{MHz}$, and effective nonlinear interaction rate $\abs{\chi_0} = 2 \pi \cross 59.2 \, \mathrm{Hz}$ such that $N \abs{\chi_0} = 2 \pi \cross 5.92 \, \mathrm{kHz}$.
We can now calculate ratios to check each of the approximations used in deriving Eq.~\eqref{H_VC_supp}, which we present in Table.~\ref{tb:approxTable}.
We find that all our approximations are satisfied by at least a factor of 10, while also satisfying $\abs{\Delta_c'} \gg \omega_g$ used in Eq.~\eqref{AlphaSimplified} by a factor $10$.
We therefore expect our simulations of Eq.~\eqref{H_VC_supp} to be a realistic model of the current vertical cavity experiment of Ref.~\cite{Luo}.
\begin{table}[ht]
    \centering
    \begin{tabular}{|c||c|c|}
    \hline
        Approximation & Inequality ($A \gg B$) & Ratio ($A / B$) \\
    \hline
        Excited state elimination & $\abs{\Delta_a} \gg \sqrt{N} \Lambda$ & 10 \\
        Cavity elimination & $\abs{\Delta_c'} \gg N \abs{\chi_0}$ & 820 \\
        Cavity elimination & $\abs{\Delta_c'} \gg \abs{U_0} \abs{\beta_0}^2$ & 42 \\
        Perturbation & $1 \gg N \abs{\epsilon} / 2$ & 39 \\
        Single momentum flips & $12 k g \tau \gg \abs{U_0} \abs{\beta_0}^2$ & 26 \\
        Unwanted pair creation & $16 \omega_r \gg N \abs{\chi_0}$ & 10 \\
    \hline
    \end{tabular}
    \caption{Table outlining the approximations assumed throughout the derivation as well as their corresponding ratios for our chosen experimental parameters.}
    \label{tb:approxTable}
\end{table}

\bibliography{references.bib}

\end{document}